\documentclass{llncs}

\usepackage{llncsdoc}
\usepackage{graphicx}

\newcommand{\keywords}[1]{\par\addvspace\baselineskip
\noindent\keywordname\enspace\ignorespaces#1}

\begin {document}
\title{Phase Selection Heuristics for Satisfiability Solvers}
\titlerunning{Phase Selection Heuristics for SAT Solvers}

\author{Jingchao Chen}
\institute{School of Informatics, Donghua University \\
2999 North Renmin Road, Songjiang District, Shanghai 201620, P. R.
China \email{chen-jc@dhu.edu.cn}}

\maketitle
\begin{abstract}
In general, a SAT Solver based on conflict-driven DPLL consists of
variable selection, phase selection, Boolean Constraint Propagation,
conflict analysis, clause learning and its database maintenance.
Optimizing any part of these components can enhance the performance
of a solver. This paper focuses on optimizing phase selection.
Although the ACE (Approximation of the Combined lookahead
Evaluation) weight is applied to a lookahead SAT solver such as
March, so far, no conflict-driven SAT solver applies successfully
the ACE weight, since computing the ACE weight is time-consuming.
Here we apply the ACE weight to partial phase selection of
conflict-driven SAT solvers. This can be seen as an improvement of
the heuristic proposed by Jeroslow-Wang (1990). We incorporate the
ACE heuristic and the existing phase selection heuristics in the new
solver MPhaseSAT, and select a phase heuristic in a way similar to
portfolio methods. Experimental results show that adding the ACE
heuristic can improve the conflict-driven solvers. Particularly on
application instances, MPhaseSAT with the ACE heuristic is
significantly better than MPhaseSAT without the ACE heuristic, and
even can solve a few SAT instances that remain unsolvable so far.

\end{abstract}

\section{Introduction}
Satisfiability (SAT) is a classic NP-complete problem, and has
applications in numerous fields such as computer aided design, data
diagnosis, EDA, logic reasoning, cryptanalysis, planning,
equivalence checking, model checking, test pattern generation etc.
To address the SAT problem, numerous state-of-the-art solvers have
been developed. This problem has been studied for a long time.
However, large real-world SAT problems remain unsolvable yet.

 Most modern SAT solvers are based on conflict-driven DPLL. The most
representative solver in this type of solvers is Precosat
\cite{PrecosatA:3,PrecosatB:4,Precosat10:5}, which won a Gold Medal
for application category at SAT 2009 competition. Generally
speaking, this type of solvers is good at solving application
instances. In this paper, we focus on a conflict-driven DPLL-type
solver like Precosat. In general, a conflict-driven DPLL-type solver
consists of variable selection, phase selection, BCP (Boolean
Constraint Propagation), conflict analysis, clause learning and its
database maintenance. Each part has various optimizing strategies.
For example, for variable selection, the corresponding optimizing
strategy is VSIDS (Variable State Independent Decaying Sum) scheme
\cite{CDCL:16}. To speed up BCP, two watched-literals scheme was
proposed. With respect to conflict analysis, a large amount of
optimizing work has been done. The research results such as firstUIP
(unique implication points), conflict clause minimization,
on-the-fly self-subsuming resolution \cite{Onfly:17}, learned clause
minimization \cite{Min:18}, have been achieved. To maintain
effectively clause learning database, recently, Audemard et al.
\cite{glue:19} introduced a Glucose style reduce strategy to remove
less important learned clauses. Although the phase heuristic is also
an important component of modern conflict-driven SAT solvers, the
literature on the phase selection is rare. To our best knowledge, up
to now, only two phase selection strategies were widely used in
conflict-driven SAT solvers. One is the phase heuristic used in RSAT
(RSAT heuristic for short) \cite{RsatHeuris:1}. The other is
Jeroslow-Wang heuristic \cite{JWheuris:2}. The basic idea of the
RSAT heuristic is to save the previous phase and assign the decision
variable to the same value when it is visited once again. The basic
idea of Jeroslow-Wang heuristic is to define variable polarity as a
phase with the maximum weight. The value of weight of a variable
depends on the number of clauses containing that variable and its
size. Although PrecoSAT \cite{Precosat10:5} gains the good
performance by integrating the RSAT heuristic and Jeroslow-Wang
heuristic, it cannot be concluded that there does not exist a better
phase heuristic.

\setlength{\parskip}{ 0.3ex} The goal of this paper is to find a new
phase selection heuristic that improves the existing phase selection
heuristics. If we can select always correctly a phase, all
satisfiable formulae will be solved in a linear number of decisions.
In theory, no perfect phase selection heuristic exists unless P=NP.
In practice, it is possible to develop a phase selection heuristic
that significantly reduces the number of decisions in some cases. To
achieve this goal, we hope to use a new phase heuristic, which uses
some information about the structure of the problem such as the
number of variables, the number of XOR clauses etc. The ACE
(Approximation of the Combined lookahead Evaluation) weight has been
applied successfully to a lookahead SAT solver such as March.
However,  so far, no conflict-driven SAT solver applies successfully
the ACE weight, since computing the ACE weight is time-consuming.
Here we apply partially the ACE weight to phase selection of
conflict-driven SAT solvers. This can be seen as an improvement of
the Jeroslow-Wang heuristic. Based on our empirical observation, the
phase selection heuristic based on ACE can enhance the ability of
solving some instances. In our solver, the ACE heuristic is only
applied under certain circumstances, mainly due to its cost. In the
case the ACE heuristic is not suited for, the other heuristics such
as Jeroslow-Wang heuristic are applied. To avoid harming the other
heuristics when applying the ACE heuristic, we define a set of
criteria for selecting the good phase heuristic in a way similar to
portfolio methods \cite
{borgsat:7,Portfolio:8,XuA:9,XuB:10,XuSovler:11} to classify
instances. Our method to classify instances uses fewer features, and
is simpler than the model-based portfolio method \cite{Portfolio:8}.
We build a new SAT solver, called MPhaseSAT, by integrating multiple
phase heuristics including the ACE heuristic and Jeroslow-Wang
heuristic.

  Empirical results show that adding the ACE heuristic can improve
our new solver MPhaseSAT. Particularly for application instances,
the improvement is significant. On this category, MPhaseSAT with the
ACE heuristic is significantly superior to MPhaseSAT without the ACE
heuristic, and even can solve a few SAT instances that remain
unsolvable so far. Although the improvement on the crafted category
is not so big, MPhaseSAT with the ACE heuristic is still a little
better than MPhaseSAT without the ACE heuristic.
\section{Phase selection heuristics}

The decision variable selection is indispensable to conflict-driven
SAT solvers. The decision heuristic used in most SAT solvers is a
more dynamic and adaptive version of the original zChaff decision
heuristic \cite {CDCL:16}. The phase selection of variable is an
inseparable step that follows the decision variable selection,
because we must assign each decision variable to a value. The
simplest phase selection heuristic is a default heuristic of
MiniSAT, in which each decision variable is always assigned to
false. To avoid work repetition caused by some independent
components, the strategy used in RSAT \cite{RsatHeuris:1,Rsat:6} is
to assign the decision variable to the same value it has been
assigned before. PrecoSAT \cite{Precosat10:5} combines RSAT
heuristic \cite{RsatHeuris:1} and Jeroslow-Wang heuristic
\cite{JWheuris:2} to select the phase of the decision variable. Its
basic idea is: when the decision variable has not been assigned yet,
Jeroslow-Wang heuristic is used. Otherwise, RSAT heuristic is used.
The basic idea of the Jeroslow-Wang heuristic is to select the phase
of a decision variable by comparing the weights of two phases. Let
$S$ define a set of CNF (Conjunctive Normal Form) clauses. This
heuristic defines the weight of $S$ as

\hskip 10mm $ W(S) = \sum\limits_{k=1}^{\infty}
\displaystyle {\frac{n_k}{2^k}} $ \\
where $n_k$ is the number of clauses of size $k$ in $S$. For a
decision variable $v$, let $S_v$ and $S_{-v}$ be the set of clauses
in which $v$ occurs positively, respectively negatively. If $W(S_v)>
W(S_{-v})$, this heuristic picks the positive phase, i.e., assign
$v$ to true. Otherwise, it picks the negative phase, i.e., assign
$v$ to false.

Jeroslow and Wang \cite{JWheuris:2} presented a simple analysis on
this heuristic, and indicated that when $W(S) < 1$, $S$ must be
satisfiable. In an intuitive sense, as long as we choose always a
literal with the maximum weight, which can yield a formula of
minimum weight, a formula is most likely to be satisfiable. However,
in real applications, this heuristic is not necessarily effective.
So far, no state-of-the-art SAT solver uses Jeroslow-Wang heuristic
to pick a decision variable. The solver PrecoSAT applies it to only
the phase selection of a decision variable, not decision variable
selection. In our experiments, we noted that in some cases
Jeroslow-Wang heuristic was efficient for the phase selection of a
decision variable, but in some cases the new heuristic given below
was more efficient.

The new heuristic uses some information about the structure of the
problem such as the number of variables, the number of XOR clauses,
etc. It defines the weight of a literal as ACE (Approximation of the
Combined lookahead Evaluation). The heuristic based on the ACE
weight here is called ACE heuristic. The concept of ACE is widely
used in lookahead SAT solvers such as March \cite {Marchhi:14},
MoRsat \cite {MoRsat:15}. The concept of ACE here is the same as
that one used in MoRsat. However, its computation is simpler than
that of ACE in March. Let the notation ${\cal F}(x=0)$ denote the
resulting formula after assigning literal $x$ to false and
performing iterative unit propagation. ${\cal F}(x=1)$ is similar.
The ACE weight of a literal $x$ is defined as
\begin{center}
 {\small
 ACE$(x,{\cal F},{\cal F}')=\sum\limits_{c \in \mathrm
{CNF}(x,{\cal F})}W_{CNF}(\mathrm {size}(c,{\cal
F}'))+\sum\limits_{c \in \mathrm {XOR}(x,{\cal F})}W_{XOR}(\mathrm
{size}(c,{\cal F}'))$}
\end{center}
\noindent where ${\cal F}'$ is either ${\cal F}(x=0)$ or ${\cal
F}(x=1)$. CNF$(x,\cal F)$ and XOR$(x, \cal F)$ are the set of CNF
clauses and the set of XOR clauses in formula $\cal F$ in which
variable $x$ occurs, respectively, and size$(c, {\cal F}')$ denotes
the length of the clause to which $c$ is reduced after an iterative
unit propagation ${\cal F}'$, and weight functions $W_{CNF}(n)$ and
$W_{XOR}(n)$ are defined as

\hskip 10mm  $W_{CNF}(n)=5^{2-n}$

\hskip 10mm  $W_{XOR}(n)=5.5 \times 0.85^n$

\noindent where $n$ is the length of the reduced clause. The
heuristic defined by ACE chooses always a phase with the maximum ACE
weight. In some sense, the ACE heuristic can be seen as an
improvement of Jeroslow-Wang heuristic defined above.

\begin{table}
\caption{ Runtime (in seconds) required by MPhaseSAT with different
phase selection heuristics to solve SAT problems.}
\begin{center}

\setlength\tabcolsep{4pt}
\begin{tabular}{l|c|c|c|c}
\hline  \hline
\multicolumn{1}{c|}{Instance} & \# var & \# clauses & ACE &
PrecoSAT \\
 &  &  & heuristic & heuristic \\
\hline cube-11-h13-unsat  & 455627 & 1367522 & 297 & 8795 \\
schup-l2s-bc56s-1-k391 & 561371 & 1778987 & 302 & 911 \\
unif2p-p0.7-v3500-c9345-S1832504551 & 3500 & 9344 & 194 & 454 \\
unif2p-p0.7-v4500-c12015-S1626790907 & 4500 & 12014 &   3679 &    5771 \\
lksat-n1000-m6860-k4-l4-s1935114289 & 1000 & 6860 & 4502 & 8360 \\
lksat-n1100-m7545-k4-l4-s310659001 & 1100 & 7545& 1952 &   2582 \\
\hline
\end{tabular}
\end{center}
\end{table}

Table\,1 shows some examples that benefit from the ACE heuristic.
PrecoSAT heuristic means the combination of Jeroslow-Wang heuristic
and RSAT heuristic, which is used in the PrecoSAT solver. All six
instances used in this experiment are unsatisfiable. The reason why
we did not choose any satisfiable instance is to rule out a lucky
solving. The first two instances in Table\,1 are from application
category in SAT 2009 competition. The middle two instances are from
random category in SAT 2007 competition. The last two instances are
from crafted category in SAT 2009 competition. The SAT solver used
is MPhaseSAT, a variant of CicleSAT \cite {CircleSAT:13}, which is
built on the top of PrecoSAT 465. As shown in Table\,1, each
category has some instances for which using the ACE heuristic was
faster than using PrecoSAT heuristics. There are also many
unsuccessful instances. For example, dated-5-15-u,
9dlx\_vliw\_at\_b\_iq2, q\_query\_3\_l48\_lambda, etc, on these
instances, using ACE the heuristic was slower on the contrary.

In theory, no matter what phase selection heuristic we use, for
unsatisfiable instances, the search efficiency should be the same,
as long as no restart occurs, the variable decision policy is the
same, and learnt clause database is infinitely extended. However, in
fact, every state-of-the-art conflict-driven solver has restart
policy and the maximum limit of learnt clause database. Therefore,
different phase selection policies perform best on different
instances. Then, for a specific instance, what is the best phase
selection? This is either an adaptive problem or a performance
prediction problem.

 In order to address better the phase selection problem, we present
the following seven phase heuristics, one or multiple ones of which
will be used for solving a SAT instance.
\begin{enumerate}
\item Jeroslow-Wang heuristic: JW heuristic for short.

\item ACE heuristic: when the search depth is smaller than 30, ACE weight is used. Otherwise, JW
weight is used.

\item JW+RSAT heuristic: a combination of Jeroslow-Wang
heuristic and RSAT heuristic. Because it is used in PrecoSAT, it is
called PrecoSAT heuristic also.

\item PrecoSAT+tail JW heuristic: within the last 20 search depths, only JW heuristics is applied,
without RSAT heuristic. In the other search depths, PrecoSAT
heuristic is applied.

\item  ACE + PrecoSAT heuristic: when the number of decisions is less than 300000, ACE heuristic is applied.
Otherwise, PrecoSAT heuristic is applied.

\item PrecoSAT+random heuristic: this is similar to CryptoMiniSat \cite {CryptoMiniSat:12} policy. In general,
the phases are calculated for each variable according to PrecoSAT
heuristic. However, sometime the phase and the decision variable are
randomly selected. Our strategy is that the decision variable is
randomly selected with the probability of 0.02, and the phase is
randomly flipped probability of 1/30.

\item Local search phase heuristic: the state of local search such as the solver TNM \cite {TNM:20} is considered as
the basis of phase selection. This heuristic chooses always a phase
in accordance with the current state of a variable local search
algorithm returns.

\end{enumerate}

Unlike JW heuristic, ACE heuristic is dynamic. That is, the ACE
 score is calculated at each step for selecting the phase of the
new decision point. Its computation cost is much more expensive than
that of JW heuristic. Therefore, as seen above, the depth where the
ACE heuristic is applied is restricted to 30. If the depth is not
limited, the ACE heuristic is not suited for large SAT instances,
since it is time-consuming.

Another important problem is that we must decide which heuristics
are suited for which SAT instances. That is, how do we classify SAT
instances? Portfolio methods are useful to how to classify SAT
instances. Xu et al. \cite {XuA:9,XuB:10} developed the SATzilla
solver, using the SAT portfolio method. Their SAT solver is built on
the pre-solver regression-based predictors of performance, and was
very successful in the 2009 competition. To improve the existing
portfolio method, Silverthorn and Miikkulainen \cite {Portfolio:8}
considered unobserved variables as latent variables, and presented
two latent class models for algorithm portfolio methods: a mixture
of multinomial distributions, and a mixture of Dirichlet compound
multinomial distributions. Based on the model-based portfolio
method, they developed the borg-sat solver \cite {borgsat:7}, which
can outperform SATzilla in all categories from their report \cite
{Portfolio:8}, and was fairly successful in SAT-Race 2010. After
investigating these portfolio methods, we found that the portfolio
methods seem not to be suited for our phase selection problem. The
latest version of SATzilla \cite {XuSovler:11} uses domain-specific
knowledge and select manually more than 90 statistics features, such
as statistics of the variable-clause graph and of DPLL probes. This
is expensive to our task during training. Borg-sat uses minimal
domain knowledge. However, when the number of assumed latent
classes, $K$, is small, the performance of borg-sat deteriorate
sharply. In our phase selection task, it is expected that the value
of $K$ is limited to be small, and the performance should still be
good. Therefore, we do not select portfolio methods to classify SAT
instances.

Instead, we develop a simple method to classify SAT instances. In
the simple method, we choose the following instance features.

\begin{enumerate}

\item Number of clauses: denoted by $\#c$.
\item Number of variables: denoted by $\#v$.
\item Ratio of clauses and variables: denoted by $\#c/\#v$.
\item Mean search depth to conflict in DPLL probing: denoted by E($\#d$)
\item Number of unfixed variables in DPLL probing: denoted by U($\#v$)
\item Number of binary clauses: denoted by $\#bin$.
\item Number of XOR clauses: denoted by $\#xor$.
\item Number of clauses of size 9 or more: denoted by L($\#c$).
\end{enumerate}

The computation cost of the above features is very cheap. Moreover,
most features are static. Unlike SATzilla, we do not make regression
analysis. Our model construction is simple. Based on the observation
on the behavior of some representative SAT instances, we classify
manually SAT instances into some categories by feature information.
Each category corresponds to a phase selection heuristic. How to map
a category of SAT instances to a phase selection heuristic? This can
be done by an adaptive algorithm, which may be described as follows.

\begin{description}
\item [(1)] When $50000<\#c<220000$, in the preprocessing phase, we use
PrecoSAT+random heuristic to determine the variable and its phase.
This heuristic is suited well for the crypto instances such as the
mizh-sha0 family. Our preprocessing is limited to 200000 decisions.

\item [(2)] We set the phase policy to PrecoSAT heuristic in the following
cases.
\begin{description}
\item [a)] $\#xor < 1000$ and $\#c > 300000$.
\item [b)] $\#xor > 2000$ and U$(\#v) < 15000$.
\item [c)] $\#c / \#v < 6$ and $\#c / 15 > \#v$ and $\#c/3 < \#bin/2$.
\item [d)] L($\#c)> 5$ and L$(\#c) < 40$.
\end{description}

 \item [(3)] We apply ACE heuristic in the following cases.
\begin{description}
\item [a)] In the preprocessing, $\#c < 18000$.
\item [b)] E$(\#d) < 30$.
\item [c)] $\#bin > 400000, \#bin > \#c/2$ and $\#v/20 > \mathrm{U}(\#v)$.
\item [d)] $\#xor > 2000, \#xor > \#c / 12$ and U$(\#v) < 15000$.
\end{description}

\item [(4)] We select PrecoSAT +JW tail heuristic in the following cases.
\begin{description}
\item [a)] $\#xor=0, \#c / \#v > 100$ and $\#v < 1500$.
\item [b)] $\#xor=0, \#c / \#v > 55$ and $\#c / \#bin < 0.9$.
\end{description}

\item [(5)] In the pre-solving, the default heuristic is set to PrecoSAT
heuristic. In the other cases, the default heuristic is set to ACE +
PrecoSAT heuristic.

 \item [(6)] For random category, we use the MoRsat \cite {MoRsat:15} solving technique and
 local search phase heuristic.

\end{description}

The above rules for classifying instances and constants are mainly
based on application instances in SAT competition 2009. Of course,
these rules can become more coarse or fine. This depends on a real
application. Based on our observation, the above rules are
sufficient for application instances in SAT 2009.

Many modern SAT solvers such as PrecoSAT do not usually support XOR
gates. Without making any modification, such solvers cannot apply
directly our ACE heuristic. However, it is easy to implement the ACE
heuristic by detecting XOR gates during preprocessing, adding
occurrence lists, and maintaining original clauses and XOR gate
database separately.

\section{Empirical evaluation}
Due to the diversity of SAT instances, in many cases, a new
heuristic is in conflict with old heuristics. However, the new
heuristic proposed in this paper is not so. By handmade refining and
classifying, we integrate this new heuristic to a SAT solver. Does
this integration harm the entire performance of the solver? This
question will be answered by our experiments.

We carried out the experiments with such a platform: Intel Core 2
Quad Q6600 CPU with speed of 2.40GHz and 2GB memory. The instances
used in the experiments are from SAT 2009 competition.  The timeouts
for solving an application instance and a crafted instance were set
to 10000 seconds and 5000 seconds, respectively. The SAT solvers
used for a comparison are PrecoSAT \cite {Precosat10:5},
CryptoMiniSat \cite {CryptoMiniSat:12} and MPhaseSAT. PrecoSAT used
here is the latest version 465 of PrecoSAT 236, which won a Gold
Medal for application category in the SAT competition 2009.
CryptoMiniSat is the winner of the Gold Medal in SAT-Race 2010.
MPhaseSAT is an improved version of CircleSAT \cite {CircleSAT:13},
which is a conflict-driven DPLL complete solver based on PrecoSAT.
MPhaseSAT consists of preprocessing, pre-solving, conflict-driven
DPLL solving and hybrid DPLL solving. The preprocessing is similar
to that of the look-ahead solver March \cite {Marchhi:14}. The
pre-solving is used to compute the feature of SAT instances in order
to select successfully the best phase heuristic. It tests run with
PrecoSAT. Once a heuristics strategy is determined, MPhaseSAT uses
that heuristics strategy to solve with an improved PrecoSAT or a
hybrid DPLL solver. For the basic principle of hybrid DPLL solving,
the reader is referred to MoRsat \cite {MoRsat:15}.

\begin{table}
\caption{ Performance of solvers on 292 application instances in SAT
2009}
\begin{center}

\setlength\tabcolsep{4pt}
\begin{tabular}{l|c|c}
\hline  \hline
\multicolumn{1}{c|}{Solver} & Instances Solved & Average time (in seconds) \\
 & & per solved instance \\
\hline
PrecoSAT 465  &  210 & 734.47 \\
CryptoMiniSat  & 212 & 780.40 \\
MPhaseSAT I   &  218 & 635.46 \\
MPhaseSAT II  &  227 & 645.02 \\
\hline
\end{tabular}
\end{center}
\end{table}

\begin{table}
\caption{ Performance of solvers on 281 crafted instances in SAT
2009}
\begin{center}

\setlength\tabcolsep{4pt}
\begin{tabular}{l|c|c}
\hline  \hline
\multicolumn{1}{c|}{Solver} & Instances Solved & Average time (in seconds) \\
 & & per solved instance \\
\hline
PrecoSAT 465  &  149 & 391.33 \\
CryptoMiniSat &  143 & 477.96 \\
MPhaseSAT I   &  176 & 360.88 \\
MPhaseSAT II  &  178 & 296.79 \\
\hline
\end{tabular}
\end{center}
\end{table}

\begin{figure}
\centering
\includegraphics[height=7.2cm]{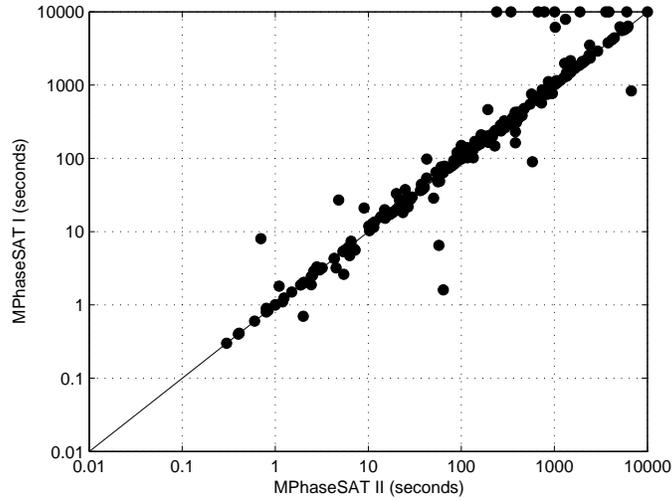}
\caption{Comparing the runtimes of MPhaseSAT I and MPhaseSAT II on
application instances from SAT 2009.} \label{singleFig}
\end{figure}

\begin{figure}
\centering
\includegraphics[height=7.2cm]{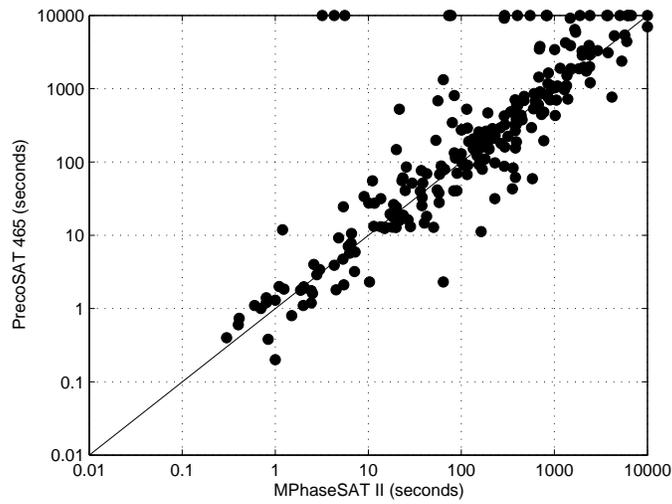}
\caption{Comparing the runtimes of MPhaseSAT II and PrecoSAT 465 on
application instances from SAT 2009.} \label{precosatFig}
\end{figure}

From our empirical results, PrecoSAT 465 outperformed indeed the
previous version PrecoSAT 236. In terms of the number of instances
solved, PrecoSAT 465 and 236 solved 210 and 204 out of 292 instances
in the application category, respectively. As shown In Table\,2,
CryptoMiniSat solved 212 instances. Except for these solved
instances, due to out of memory, 5 instances were not solved by
CryptoMiniSat. If memory is sufficient, CryptoMiniSat could solve
217 instances. To measure efficiency of the new phase heuristic, we
divided MPhaseSAT into two versions. One is the version with
multiple phase heuristics, which integrates the new phase heuristic.
The other is the version with single phase heuristic, which uses
only PrecoSAT heuristic. MPhaseSAT with single phase heuristic is
MPhaseSAT I for short. MPhaseSAT with multiple phase heuristics is
MPhaseSAT II for short. As shown in Table\,2, for the application
category, MPhaseSAT I and II solved 218 and 227 out of 292
instances, respectively. This result reveals that adding the new
phase heuristic can improve significantly the performance of the
solver. The reason why MPhaseSAT I without ACE heuristic was also
better than CryptoMiniSat is because it benefits from the
combination of pre-solving technique and hybrid DPLL technique.
Because the techniques are not the focus of this paper, we here
omitted their implementation details. The virtual best solver in the
SAT 2009 competition, which is defined as a theoretical solver which
returns the best answer provided by one of all the submitted
solvers, solved 229 application instances, 225 out of which were
solved by MPhaseSAT II. Two instances gss-24-s100 and
eq.atree.braun.12 are not included by the 229 instances, which
cannot be solved in a reasonable time by any solver so far.
Gss-24-s100 was solved by both MPhaseSAT I and II. But
eq.atree.braun.12 was solved by only MPhaseSAT II. This is due to
the application of the ACE phase heuristic. From this empirical
results, it is easy to see that MPhaseSAT II approaches very much
the performance of the virtual best solver.

We tested also the effectiveness of the ACE heuristic on the crafted
instances from SAT 2009. Because the rules for classifying instances
are determined mainly according to the behaviour of our solver on
the application category, the ACE heuristic was not very effective
on the crafted category in this experiment. As shown in Table\,3,
MPhaseSAT II is a little better than MPhaseSAT I, since they solved
178 and 176 out of 281 instances, respectively. However, they are
much better than the other two solvers. CryptoMiniSat and PrecoSAT
solved 143 and 149 out of 281 instances. Notice, the solver clasp,
the champion of this category, solved 156 instances. Therefore, both
CryptoMiniSat and PrecoSAT are not good at the crafted category.
Nevertheless, the excellent behaviour of MPhaseSAT on the crafted
category is mainly due to the other technique development including
the new at-most-one encoding technique \cite {chenAMO:21}, not the
phase heuristic given here.

Figures 1 and 2 show a log-log scatter plot comparing the runtimes
of \linebreak
 MPhaseSAT I and MPhaseSAT II, and the runtimes of MPhaseSAT II
and PrecoSAT 465, respectively. The instances in Figures 1 and 2 are
from the application category at SAT 2009. The climax (10000,10000)
means that the instances on that point were not solved by any of two
solvers. Whether in Figure\,1 or in Figure\,2, except for the points
that were not solved by MPhaseSAT I nor PrecoSAT 465, most of points
are centralised at the nearby diagonal. This demonstrates that
adding the new phase heuristic has no strong impact on solving the
instances that is not suited for the new phase heuristic. In fact,
this is verified by the average runtime per solved instance shown in
Table\,2, since the difference between the average runtimes of
MPhaseSAT I and MPhaseSAT II is very small.

If the performance of a single solver is equivalent to that of the
virtual best solver, it means than all the optmizing techniques can
be integrated in a SAT solver. Nevertheless, now it seems impossible
to do this. The reason why MPhaseSAT on the application category
approached the virtual best solver is because the training set for
establishing the rules were also application instances. Crafted
instances were not used for the training set. On the crafted
category, the performance of MPhaseSAT did not approach that of the
virtual best solver. This is reflected by the fact the virtual best
solver solved 187 instances in the crafted category at SAT 2009, 166
out of which were solved by MPhaseSAT II. The other 12 instances
solved by MPhaseSAT II remains unsolvable so far.

Like the other conflict-driven solvers, MPhaseSAT is not good at the
random category. So here we do not discuss the performance of
MPhaseSAT on the random category.

\section {Conclusions}
By integrating the ACE phase selection heuristic and the existing
heuristics such as Jeroslow-Wang heuristic and RSAT heuristic, we
have built a new SAT solver called MPhaseSAT. Surprisingly,
MPhaseSAT can not only outperform PrecoSAT, the champion of the
application category in SAT 2009, and CryptoMiniSat, the champion of
SAT-Race 2010, but also approaches the virtual best solver in terms
of the solving ability. Moreover, it solved a few SAT instances that
remain unsolvable so far.

MPhaseSAT is built on the top of the known better solver. Hence, its
performance depends heavily on that of the known solvers. Because of
the limit of the known technique, perhaps, the usefulness of the new
phase selection heuristic has not yet been tapped sufficiently. To
obtain a great breakthrough, we must devise a new solving technique
different from the existing one.

In this paper, a SAT problem is envisioned as a portfolio. During
the entire solving process, a SAT problem uses only one phase
heuristic. If a problem consists of multiple independent
sub-problems, and each sub-problem has a different portfolio, the
classifier of SAT instances given in this paper will fail, since it
cannot switch between different sub-problems. Therefore, the
adaptive switching technique among different phase heuristics needs
to be studied further.

The ACE heuristic is time-consuming, and much more expensive than
the existing phase selection heuristics. It is a subject that's
worth studying how to simplify the ACE heuristic and reduce its
computation cost.

Whether in theory or in practice, we believe that the phase
heuristics known so far are not certainly the best, and there exists
certainly a better phase heuristics. It is a very valuable research
topic how to find out a better phase heuristics.

\bibliographystyle{splncs}
\bibliography{chenCP11}

\end{document}